\documentclass[aps,prb,twocolumn,groupedaddress,showpacs]{revtex4}
\usepackage{epsfig}
\usepackage{graphicx}

\newcommand{\up}{\uparrow}
\newcommand{\dn}{\downarrow}

\renewcommand{\Re}{{\rm Re}}
\renewcommand{\Im}{{\rm Im}}

\begin{document}

\title{Role of a spin-flip scatterer in a magnetized Luttinger liquid}
\author{M.A.N Ara\'ujo$^{a,b}$,  J. Berakdar$^{c}$, V.K. Dugaev$^{b,d}$ and V. R. Vieira$^{b,e}$}
\affiliation{$^{a}$Departamento de F\'{\i}sica, Universidade de
\'Evora, \'Evora, Portugal} \affiliation{$^b$CFIF, Instituto
Superior T\'ecnico, Lisbon, Portugal} \affiliation{$^c$Insitut
f\"ur Physik, Martin-Luther-Universit\"at Halle-Wittenberg,
Heinrich-Damerow-Str.~4, 06120 Halle, Germany}
\affiliation{$^d$Department of Mathematics and Applied Physics,
Rzesz\'ow University of Technology, Al. Powsta\'nc\'ow Warszawy 6,
35-959 Rzesz\'ow, Poland} \affiliation{$^{e}$Departamento de
F\'isica, Instituto Superior T\'ecnico, Lisbon, Portugal}

\begin{abstract}
We study the spin-dependent  scattering of charge carriers in a
magnetized one dimensional  Luttinger liquid from a localized
non-homogeneous magnetic field, which might be brought about by the
stray field of  magnetic tip near a uniform liquid, 
or by a transverse domain wall (DW) between two oppositely magnetized liquids.
From a  renormalization group treatment of the  electron
interactions we deduce scaling equations for the transmission and
reflection amplitudes as the bandwidth is
progressively reduced to an energy scale set by the temperature.
The repulsive interactions dictate two
possible zero temperature insulator fixed points: one in which
electrons are reflected in the same spin channel and another where
the electron spin is reversed upon reflection. In the latter case,
a finite spin current emerges in the absence of
a charge current at zero temperature  and the Friedel oscillations 
form a transverse spiraling spin density. Adding a purely
potential scattering term has no effect on the fixed points of a
uniformly magnetized liquid. 
For a  DW we find that the introduction of potential scattering 
stabilizes the spin-flip insulator phase even if the single-particle 
spin-flip scattering produced by the DW is arbitrarily weak. 
The potential can be induced externally,
e.g. by a local gate voltage or a constriction, providing a means for
controlling the transport properties of the wire.
\end{abstract}


\maketitle


\section{Introduction}
\label{intro}

 Transport properties of magnetic nanowires are currently in the
 focus of intense research in view of possible applications in
 magneto-electronic devices \cite{wolf01,garcia99,chopra02,ruster03,hayashi07}.
Particularly interesting are  magnetic nanowires with a localized
topological magnetic disorder which acts as a spin-dependent
scatterer of charge carriers. Examples of this situations are
magnetized wires with  domain walls or magnetic microvortices that
can be well controlled externally by a magnetic field or by an
electric current. 

Here, we address the problem of scattering of interacting electrons 
from a single defect in a magnetic nanowire,
assuming that the nanowire is thin enough so that the electron energy spectrum is
one-dimensional. A similar problem in the nonmagnetic case attracted a lot of attention
in the past\cite{kane92,glazman} because of the key role of electron-electron interactions
leading to  substantial renormalization of the scattering amplitudes. 

Recently\cite{araujo06}, we  treated the problem of the spin
dependent scattering in a short transverse domain wall (DW) separating two
oppositely magnetized regions in a 1D wire. In particular, we
considered the case of a DW whose extension  is comparable to the
carriers' de Broglie wavelength, in which case  the influence of
scattering and interactions is particularly strong. On the other
hand, this limitation implies  severe demands on an experimental
realization restricting possible systems to   magnetic
semiconductor nanowires, such as (Ga,Mn)As \cite{comment}.  
A further complication is brought about by  magnetic impurities in 
the semiconductor that cause scattering in the region of the domain 
wall\cite{falloon06}. If the DW is pinned by a constriction, 
the constriction itself behaves as a pure potential scatterer.

There is  yet another possibility for
inducing localized magnetic imperfections in a wire: the stray
magnetic field of a nearby tip of a spin-polarized scanning
tunneling microscope (STM) in the T-shape geometry\cite{schlikum} 
illustrated in Fig.\ref{system}. In
this case, the wire itself may be magnetically homogeneous and 
free of impurities.

For magnetic nanowires the scattering involves  at least two channels and
different components of the transmission amplitudes may well be
renormalized in  different ways. As shown below, this results in
some interesting effects such as 
a spin-flip insulator state. The possibility of these effects has
been demonstrated recently for scattering from a magnetic transverse 
domain wall \cite{Dugaev_jpa,araujo06}. In the case of a homogeneous
magnetization of the wire, possible types of defects can be
classified as purely potential defects (i.e., a local perturbation
that  affects equally the spin up and down electrons), local
variation of the magnitude of the magnetization, and a local
variation of the magnetization direction. The first two cases are
straightforwardly treated since they can be  analyzed in terms of
potential scattering in each of the spin channels. The
problem tackled here is  the case of a local variation of
the magnetization direction, when both spin-flip and non-spin-flip
scattering amplitudes are renormalized, by electron interactions, 
in different ways.

Technically, we need to address the scattering of electrons from a
localized transverse magnetic field. 
The single-particle scattering (either pure potential or spin dependent) 
can be  treated exactly.  
We then address the role of electron interactions perturbatively, 
to first  order, resulting in the well-known logarithmic divergences. 
The divergent terms are then circumvented by   a poor man's scaling
approach that yields  a set of renormalization equations for the
scattering amplitudes. 

The two problems have similar fixed point  solutions at zero
temperature ($T=0$).  Considering repulsive spin dependent
interactions, two types of insulator may arise, depending on the
interaction parameters: the electrons may be 100\% reflected with
or without spin reversal. In the former case  there is a spin
current without charge current and the spin current exerts a
torque on the magnetic tip.
For the case of a DW we find the counter-intuitive effect that a 
non-spin-flip  insulator phase is obtained if the DW's transverse field is weak. 
But by adding a purely potential scattering term, 
one is able to drive the DW to the spin-flip insulator phase.
The potential itself can be externally imparted, e.g.  by
a constriction or by a local gate voltage, much in the same way
as demonstrated by recent experiments \cite{Topinka}.

Section \ref{model} introduces the electronic Hamiltonian for the 
problems depicted in Figure \ref{system} and the single-particle solutions. 
Section \ref{scalings} gives the renormalization group equations 
for the scattering amplitudes in each problem, which are a 
consequence of the electron interactions. 
Section \ref{theend} contains a discussion and summary.

\section{Model Hamiltonian}
\label{model}

\begin{figure}[ht]
\begin{center}
\epsfxsize=8cm \epsfbox{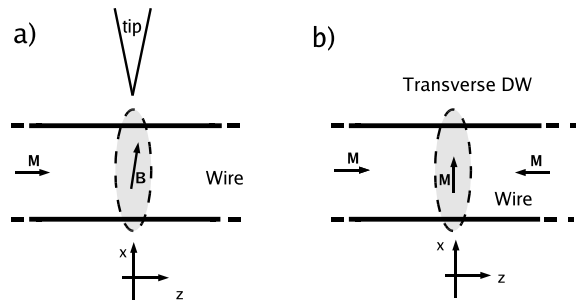}
\end{center}
\caption{(a) Magnetic tip produces a nearly transverse magnetic
field in a localized region of a uniformly magnetized metallic
wire; (b) a transverse DW in the wire also acts as a transverse
field.} \label{system}
\end{figure}

We consider a ferromagnetic metallic wire close to a magnetic tip
as shown in Figure \ref{system}(a). 
The latter produces an effective magnetic field which acts 
as a spin-flip scatterer in a localized region of the wire.
The wire defines the easy ($\hat z$) axis with uniform magnetization ${\bf M} =  M\hat z$.
The effective  magnetic field $B_\perp(z)\hat x +B_{||}(z)\hat z$ due to the 
influence of the tip which is placed nearly perpendicular to  the wire affects a
small region near $z=0$. The conduction electron spin in the wire
is Zeeman coupled to this magnetic field. We treat the conduction
electrons as one-dimensional and write the single particle
Hamiltonian as:
\begin{equation}
\hat H_0 = -\frac{\hbar^2}{2m} \frac{d^2}{dz^2} + \hbar V
\delta(z) -J M_0\hat \sigma_z - \hbar \lambda   \delta(z)\hat
\sigma_x
 - \hbar \lambda'   \delta(z)\hat \sigma_z\,.
\label{tipmodel}
\end{equation}
The terms $\hbar \lambda^{(')}   \delta(z)\hat \sigma_x$,  describe spin
scattering produced by the  the magnetic tip:
\begin{eqnarray*}
\hbar\lambda &=& \mu\int_{-\infty}^{\infty}B_\perp(z)dz\,, \\
\hbar\lambda' &=& \mu\int_{-\infty}^{\infty}B_{||}(z)dz\,,
\end{eqnarray*}
and $\mu$ is the electron's magnetic moment. We also allow for a
purely  potential scattering term, $V$, that may be present.

In the case where the wire has two opposite ferromagnetic domains
separated by a thin transverse DW, as shown in Figure
\ref{system}(b),
 the single particle Hamiltonian is:
\begin{equation}
\hat H_{DW} = -\frac{\hbar^2}{2m} \frac{d^2}{dz^2} + \hbar V
\delta(z) -J M(z)\hat \sigma_z - \hbar\bar \lambda   \delta(z)\hat
\sigma_x\,. \label{DWmodel}
\end{equation}
where the term $\hbar \bar \lambda   \delta(z)\hat \sigma_x$ now
produces spin-flip scattering due to the $x$ component of the
magnetization in the DW:
$$
\hbar\bar \lambda = -J\int_{-\infty}^{\infty}M_x(z)dz\,
$$
and the longitudinal magnetization $M(z)=-M_0$ for $z<0$ and
 $M(z)=M_0$ for $z>0$. In this case the longitudinal magnetization produces a Zeeman
potential step which scatters electrons more strongly as compared
to the magnetic tip problem above.

In both models, a single spin-majority electron may be transmitted
either preserving or reversing its spin with amplitudes $t_\up$ or
$t_\up'$, respectively. A spin minority electron has transmission
amplitudes $t_\dn$ or $t_\dn'$. The reflection amplitudes in the
same (opposite) spin channel are denoted by $r_\up$ ($r_\up'$) for
spin majority  and  spin minority electrons, respectively.

We wish to consider the effect of electron interactions on the
scattering amplitudes. The interactions can be described by the
g-ology model\cite{Solyom}:
\begin{eqnarray}
\hat H_{int} &=& g_{1,\alpha,\beta}\int
\frac{dk_1dq}{(2\pi)^2}\hat a^\dagger_{k_1,\alpha} \hat
b^\dagger_{k_2,\beta}
 \hat a_{k_2+q,\beta} \hat b_{k_1-q,\alpha}\nonumber\\
&+&  g_{2,\alpha,\beta}\int \frac{dk_1dq}{(2\pi)^2}\hat
a^\dagger_{k_1,\alpha} \hat b^\dagger_{k_2,\beta}
 \hat b_{k_2+q,\beta} \hat a_{k_1-q,\alpha} \,.
\label{hint}
\end{eqnarray}
The couplings $g_1$ and $g_2$ describe back and forward scattering
processes between opposite moving electrons, respectively, and are
positive if the interactions are repulsive. Because the Fermi
momentum depends on spin,  we allow for the dependence of  $g$ on
the spins of the interacting particles. We therefore distinguish
between $g_{1\up}$, $g_{1\dn}$  $g_{1\perp}$ and  $g_{2\up}$,
$g_{2\dn}$, $g_{2\perp}$. The $g_2$ processes imply zero momentum
transfer whereas the $g_1$ processes involve a $2k_F$ momentum
transfer. We therefore expect $g_2$ larger than $g_1$ for a finite
spatial range of the repulsive interactions.

Our treatment of the interactions follows the same method invented
in Ref\cite{glazman} and developed in Ref \cite{araujo06}: the
corrections to the scattering amplitudes  are calculated to first
order in the perturbation $\hat H_{int}$, leading to logarithmic
divergent terms if the electron is near the Fermi level. Using  a
poor man's scaling procedure, the divergent terms lead to  a
renormalization of the scattering amplitudes as the bandwidth, $D$,
is progressively reduced from its initial value, $D_0$, to $D=T$.
The fixed points are attained at temperature $T=0$.

\section{Scaling equations}
\label{scalings}

We write down the renormalization group (RG) equations for the
scattering amplitudes, in each problem, using  the variable $\xi =
\log\left(D/D_0\right)$ which will be integrated from $0$ to
$\log\left(T/D_0\right)$, corresponding to the fact that the
bandwidth is progressively reduced from $D_0$ to $T$. We also
define:
\begin{eqnarray}
g_\sigma = \frac{g_{2\sigma}-g_{1\sigma}}{4hv_\sigma}\,, \qquad
g_\perp = \frac{g_{2\perp}}{2h(v_\dn + v_\up)}\,.
\end{eqnarray}
where $v_\up$ ($v_\dn$) denotes the Fermi velocity of spin
majority (minority) particles.

The RG equations result from a first order perturbative treatment of the interactions.
The total scattering (transmission or reflection) amplitude 
is the sum over all virtual scattering processes involving 
the electron-electron interaction once\cite{araujo06}. The latter occurs
in the form of  Bragg back-scattering caused by the $2k_F$ Hartree-Fock potential
caused by the Fermi sea Friedel oscillations.  
So, the simplest process is just one Bragg reflection 
produced by the Friedel oscillation.
The other possible process involve three virtual scattering events: the electron 
first collides with the  barrier, then with the Friedel oscillation and 
finally with the barrier again.

\subsection{Thin DW}

A detailed derivation of the RG equations for this case has been
given elsewhere\cite{araujo06} and we merely reproduce them here:

\begin{eqnarray}
\frac{d t_\sigma}{d\xi} &=&
g_\sigma\left( \  |r_\sigma|^2  t_\sigma + r_\sigma^*  r_\sigma' t_\sigma'\  \right)\nonumber\\
&+&
 g_{-\sigma} \left( \   |r_{-\sigma}|^2 t_\sigma + r_{-\sigma}^* r_\sigma' t_{-\sigma}'\ \right)  \nonumber\\
&+& g_\perp \left( 2 r_{-\sigma}'^*  r_\sigma' t_\sigma\  +\
r_\sigma'^*  r_{-\sigma} t_\sigma'\  +\ r_\sigma'^* r_\sigma
t_{-\sigma}'\   \right)\,, \label{tupdif}
\end{eqnarray}
\begin{eqnarray}
\frac{d t_\sigma'}{d\xi} &=&
2g_\sigma  | r_\sigma|^2 t_\sigma'\ +\  2g_{-\sigma}   r_{-\sigma}^*  r_{-\sigma}' t_\sigma\nonumber\\
&+& 2 g_\perp \left( r_{-\sigma}'^* r_\sigma t_\sigma\ + \
r_\sigma'^* r_{-\sigma}' t_\sigma'\ \right)\,, \label{tpupdif}
\end{eqnarray}


\begin{eqnarray}
\frac{ d r_\sigma}{d\xi} &=&
 g_\sigma \left( |r_\sigma|^2  r_\sigma\  +\  r_\sigma^* t_\sigma'^2 -r_\sigma \right)\nonumber\\ &+&
g_{-\sigma} \left(  r_{-\sigma}^*  t_\sigma  t_{-\sigma}\  +r_{-\sigma}^* r_{-\sigma}'  r_\sigma'\  \right)\nonumber\\
&+& 2 g_\perp \left( \  r_\sigma'^* r_{-\sigma}' r_\sigma\
 +\
r_\sigma'^* t_{-\sigma} t_\sigma'\  \right)\,, \label{ruplogs}
\end{eqnarray}


\begin{eqnarray}
\frac{ d r_\sigma'}{d\xi} &=&
 g_\sigma \left( |r_\sigma|^2  r_\sigma'\  +\ r_\sigma^* t_\sigma't_\sigma\ \right)\nonumber\\ &+&
g_{-\sigma} \left( r_{-\sigma}^*  t_\sigma  t_{-\sigma}'\  + r_{-\sigma}^* r_{-\sigma}  r_\sigma'\  \right) \nonumber\\
&+& g_\perp \left( \  r_{-\sigma}'^* r_{-\sigma} r_\sigma\ +
r_{-\sigma}'^* r_\sigma'^2\  +\ r_\sigma'^* t_{-\sigma}'
t_\sigma'\ \right. \nonumber\\ && \hspace{0.6cm} +\ r_{-\sigma}'^*
t_\sigma^2  \left. \ - \ r_\sigma' \    \right)\,. \label{rupup}
\end{eqnarray}
The Wronskian theorem\cite{araujo06,messiah} for the scattering
problem establishes the following relations between the scattering
amplitudes:
\begin{eqnarray}
|r_\sigma|^2 + |t_\sigma'|^2 + \frac{v_{-\sigma}}{v_\sigma} \left(
|r_\sigma'|^2 + |t_\sigma|^2 \right) =1\,, &&
\label{DW1}\\
r_\sigma^* r_{-\sigma}' +\ t_\sigma'^*t_{-\sigma} +\
\frac{v_{-\sigma}}{v_\sigma} \left( r_\sigma'^*r_{-\sigma}+\
t_\sigma^* t_{-\sigma}' \right) = 0\,, && \label{DW2} \\
r_\sigma^* t_{-\sigma} +\ t_\sigma'^*r_{-\sigma}' +\
\frac{v_{-\sigma}}{v_\sigma} \left( r_\sigma'^*t_{-\sigma}'+\
t_\sigma^* r_{-\sigma} \right) = 0\,, && \label{DW3} \\
\Re[t_\sigma'^*r_\sigma]\  +\ \frac{v_{-\sigma}}{v_\sigma} \Re[r_\sigma'^*t_\sigma] = 0\,, && \label{DW4} \\
v_\sigma t_{-\sigma} = v_{-\sigma} t_\sigma \,, \hspace{0.5cm}
v_\sigma r_{-\sigma}' = v_{-\sigma} r_\sigma'\,. && \label{DW5}
\end{eqnarray}

The initial non-interacting values ($\xi=0$) of the scattering
amplitudes near the Fermi level are obtained from the Hamiltonian
(\ref{DWmodel}):
\begin{eqnarray}
t_\sigma &=& \frac{2v_\sigma \left( v_\sigma+v_{-\sigma} + 2iV
\right) }
{ \left(v_\sigma+v_{-\sigma} + 2iV\right)^2 + 4\bar{\lambda}^2} = r_\sigma + 1\,,\label{tup}\\
t_\sigma' &=& \frac{4i\bar{\lambda} v_\sigma}{\left(v_\sigma+v_{-\sigma}
+ 2iV\right)^2 + 4\bar{\lambda}^2} = r_\sigma'\,.\label{tupl}
\end{eqnarray}
In the absence\cite{araujo06} of potential scattering ($V=0$),
$t_\sigma(\xi)$ and $r_\sigma(\xi)$ are real while
$t_\sigma'(\xi)$ and $r_\sigma'(\xi)$ are pure imaginary. In this
case, two types of insulator fixed points were
found\cite{araujo06}: the ordinary insulator with
$t_\sigma=t_\sigma'=r_\sigma'=0$, $|r_\sigma|=1$, which is
attained if $g_\perp < \left( g_\up+g_\dn \right)/2$; the
"spin-flip reflector" with  $t_\sigma=t_\sigma'=r_\sigma=0$,
$|r_\sigma'|=\sqrt{v_\up/v_\dn}$, which is attained for $g_\perp >
\left( g_\up+g_\dn \right)/2$ {\it and} not small $\bar{\lambda}/v_\up$.

\begin{figure}[ht]
\begin{center}
\epsfxsize=8cm \epsfbox{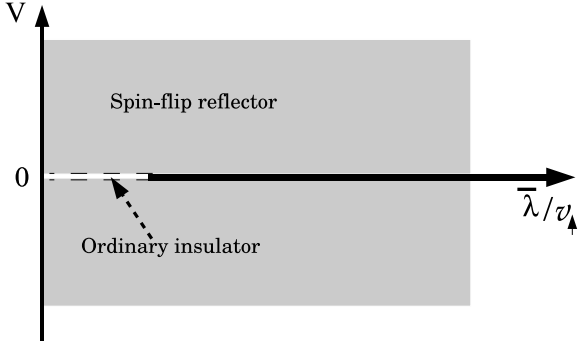}
\end{center}
\caption{Zero temperature phase diagram of a DW for $g_\perp >
\left( g_\up+g_\dn \right)/2$. Electron reflection is accompanied
with spin reversal everywhere except at small $\bar{\lambda}$ and
$V=0$.} \label{fases}
\end{figure}

Our aim here is to analyze the effect of the purely potential term
$V$. For small $\bar{\lambda}/v_\up$ and $V=0$, the DW flows to the zero
temperature fixed point with $t= t'=r'=0$ and  $r_\up=-r_\dn=1$.
The particles are reflected in the same spin channel because the
spin-flip term $\bar{\lambda}/v_\up$ is not strong enough and the
Zeeman potential step $-J M(z)\hat \sigma_z$ in (\ref{DWmodel}) is
dominant. For finite  potential scattering, $V$, all the
amplitudes become complex. We may study the RG flow near this
fixed point by linearizing equation (\ref{rupup}) for $\sigma=\up$
in the small quantity $r_\up'$:

\begin{eqnarray}
\frac{ d r_\up'}{d\xi} &\approx& \left( g_\up + g_\dn- g_
\perp\right) r_\up'  + g_ \perp  r_\up r_\dn  r_\up'^*    \,,
\label{rlupaprox}
\end{eqnarray}
which, for $ r_\up r_\dn=-1$, gives the scaling of the real and
imaginary parts as
\begin{eqnarray}
\Re[r_\up'] \propto e^{\left( g_\up + g_\dn - 2g_\perp\right)\xi}
\,, \hspace{0.4cm} \Im [r_\up'] \propto e^{\left( g_\up +
g_\dn\right)\xi} \,,
\end{eqnarray}
implying that $\Im[r_\up']$ scales to zero  as $T\rightarrow 0$
(or $\xi\rightarrow -\infty$), while
 $\Re [r_\up']$ is relevant for $g_\perp > \left( g_\up+g_\dn \right)/2$
and irrelevant if $g_\perp < \left( g_\up+g_\dn \right)/2$. Since
the real part of $r_\up'$ is due to the finiteness of $V$, we see
that the potential scattering will enhance the spin-flip
reflection processes if $g_\perp > \left( g_\up+g_\dn \right)/2$.
For such interactions, the $r_\sigma'=0$ fixed point is in fact
{\it unstable to any small} $V$ and the system will flow to the
other fixed point with $r_\sigma=0$ and $|r_\up'| =
\sqrt{v_\up/v_\dn}$. This is shown in Figure \ref{fases}.


Considering stronger $\bar{\lambda}/v_\up$, $g_\perp < \left( g_\up+g_\dn \right)/2$ and $V=0$ 
the zero temperature fixed
point has $t= t'=r'=0$ and  $r_\up=r_\dn=1$. Near this fixed
point, equation (\ref{rlupaprox}) yields:
\begin{eqnarray}
\Re[r_\up'] \propto e^{\left( g_\up + g_\dn\right)\xi} \,,
\hspace{0.4cm} \Im [r_\up'] \propto e^{\left( g_\up +
g_\dn-2g_\perp\right)\xi} \,.
\end{eqnarray}
The potential $V$, responsible for $\Re[r_\up']$, is, therefore,
irrelevant. This fixed point is stable for interactions obeying
$g_\perp < \left( g_\up+g_\dn \right)/2$ and unstable otherwise,
thereby taking the system to the spin-flip reflector fixed point.
For general finite $V$, the interactions determine whether the
fixed point has $|r|=1$, for $g_\perp < \left( g_\up+g_\dn
\right)/2$, or $|r_\up'| = \sqrt{v_\up/v_\dn}$ when $g_\perp >
\left( g_\up+g_\dn \right)/2$.

We can provide a physical explanation for the behavior of the DW 
shown in Figure \ref{fases} for small $\bar{\lambda}$. 
If the DW's transverse field ($\bar{\lambda}$) is zero, 
the renormalized reflection amplitudes $r_\up$ and $r_\dn$ tend to 1 and -1, respectively.
There is no reflection with spin flip as there is no spin-flip scattering term.
Now consider an incident electron from the left with spin up. 
What is the probability amplitude of a spin-flip reflection for small $\bar{\lambda}$?
There are two processes to first order both in the interactions $g$ and $r_\up'$:
($i$) the electron is Bragg reflected back, in the $\dn$-spin channel,  
by the Friedel oscillations of spin-down electrons with  a
probability amplitude $g_\perp r_\up'$; 
($ii$) the electron is first reflected by the DW with amplitude $r_\up$, 
then it is spin-flip back-scattered by the  $\up$-spin Friedel oscillation with an
amplitude  $g_\perp r_\up'^*$, and finally reflected by the DW with amplitude $r_\dn$. 
Since $r_\up r_\dn=-1$ and  $r_\sigma'$ is pure imaginary,  process (ii) has exactly 
symmetrical amplitude to process ($i$). So, for $V=0$, the two processes interfere destructively, 
resulting in the absence of spin-flip reflection.
If $V \neq 0$, the amplitudes in processes ($i$) and ($ii$) acquire different phases
and  no longer cancel each other. In this case there is a reflection with spin reversal, 
with amplitude proportional to $g_\perp$, and if $g_\perp > \frac 1 2 \left(g_\up +
g_\dn\right)$ the system will flow to the spin flip reflector phase.
A strong $\bar{\lambda}$ leads to the fixed point with $r_\up=r_\dn=1$ and is this case
there is no destructive interference.

\subsection{Magnetic tip}
The derivation of the RG equations for the magnetic tip problem
follows exactly the same method as for the DW. In fact, the
scaling equations for this case can simply be written down by
inspection of the same Feynman diagrams:



\begin{eqnarray}
\frac{d t_\sigma}{d\xi} &=& 2 g_\sigma |r_\sigma|^2  t_\sigma
+ 2 g_{-\sigma}  r_{-\sigma}^* r_\sigma' t_{-\sigma}'\nonumber\\
&+&  2 g_\perp \left( r_{-\sigma}'^*  r_\sigma' t_\sigma\  +\
r_{-\sigma}'^* r_\sigma t_\sigma'\ \right)\,, \label{tuptip}
\end{eqnarray}


\begin{eqnarray}
\frac{d t_\sigma'}{d\xi} &=&
g_\sigma \left(  |r_\sigma|^2 t_\sigma' + r_\sigma^* r_\sigma' t_\sigma \ \right)\nonumber\\
&+& g_{-\sigma}  \left( r_{-\sigma}^* r_\sigma' t_{-\sigma} + |r_{-\sigma}|^2 t_\sigma' \ \right) \nonumber\\
&+& g_\perp \left( 2 r_{-\sigma}'^* r_\sigma' t_\sigma' \ + \
r_\sigma'^* r_\sigma t_{-\sigma}\ + r_\sigma'^*  r_{-\sigma}
t_\sigma \right)\,, \label{tluptip}
\end{eqnarray}


\begin{eqnarray}
\frac{ d r_\sigma}{d\xi} &=&
 g_\sigma \left(  |r_\sigma|^2  r_\sigma\  +\  r_\sigma^* t_\sigma^2 -r_\sigma \right)\nonumber\\   &+&
g_{-\sigma} \left(  r_{-\sigma}^*  t_\sigma'  t_{-\sigma}'\  +\ r_{-\sigma}^* r_{-\sigma}'  r_\sigma'\  \right)\nonumber\\
&+& 2g_\perp \left( \  r_\sigma'^* r_{-\sigma}' r_\sigma\ +
r_\sigma'^* t_\sigma t_{-\sigma}'\ \right)\,, \label{ruptip}
\end{eqnarray}


\begin{eqnarray}
\frac{ d r_\sigma'}{d\xi} &=&
 g_\sigma \left( |r_\sigma|^2  r_\sigma'\  +\ r_\sigma^* t_\sigma't_\sigma\ \right) \nonumber\\ &+&
g_{-\sigma} \left( |r_{-\sigma}|^2  r_\sigma'    + r_{-\sigma}^* t_{-\sigma}  t_\sigma'\  \right) \nonumber\\
&+& g_\perp \left( \  r_\sigma'^* r_{-\sigma} r_\sigma\ +
r_{-\sigma}'^* r_\sigma'^2 \ + \ r_\sigma'^* t_{-\sigma} t_\sigma
\right. \nonumber\\   && \hspace{0.6cm} +\ r_\sigma'^* t_\sigma'
t_{-\sigma}'\left.  - r_\sigma'    \right)\,. \label{rluptip}
\end{eqnarray}
The relations between the scattering amplitudes which follow from
the Wronskian theorem are:
\begin{eqnarray}
|r_\sigma|^2 + |t_\sigma|^2 + \frac{v_{-\sigma}}{v_\sigma} \left(
|r_\sigma'|^2 + |t_\sigma'|^2 \right) =1\,, &&
\label{W1}\\
r_\sigma^* r_{-\sigma}' +\ t_\sigma^*t_{-\sigma}' +\
\frac{v_{-\sigma}}{v_\sigma} \left( r_\sigma'^*r_{-\sigma}+\
t_\sigma'^* t_{-\sigma} \right) = 0\,, && \label{W2} \\
r_\sigma^*t_{-\sigma}'+ t_\sigma^*r_{-\sigma}' +\
\frac{v_{-\sigma}}{v_\sigma} \left( r_\sigma'^*t_{-\sigma}+\
t_\sigma'^* r_{-\sigma}\right) = 0\,, && \label{W4} \\
\Re[r_\sigma^*t_\sigma]\ +\ \frac{v_{-\sigma}}{v_\sigma} \Re[ r_\sigma'^*t_\sigma' ] = 0\,, && \label{W3} \\
v_\sigma t_{-\sigma}' = v_{-\sigma} t_\sigma' \,, \hspace{0.5cm}
v_\sigma r_{-\sigma}' = v_{-\sigma} r_\sigma'\,. && \label{W5}
\end{eqnarray}

The initial scattering amplitudes are the (non-interacting)
solutions to (\ref{tipmodel}):
\begin{eqnarray}
t_\sigma &=& \frac{v_\sigma \left[ v_{-\sigma} +i
\left(V-\lambda'\right)\right] } {\left[v_\sigma+ i
\left(V+\lambda'\right)\right]\cdot\left[v_{-\sigma} + i
\left(V-\lambda'\right)\right] + \lambda^2}
\nonumber\\  &=& r_\sigma + 1\,, \\
t_\sigma' &=& \frac{i\lambda v_\sigma}{\left[v_\sigma+ i \left(V+\lambda'\right)\right]\cdot\left[v_{-\sigma}
 + i \left(V-\lambda'\right)\right] + \lambda^2} = r_\sigma'\,.\nonumber\\
\end{eqnarray}

In this problem, the zero temperature fixed point is entirely
determined by the interactions: if $g_\perp < \left( g_\up+g_\dn
\right)/2$ the system will flow to $r'=0$, $|r|=1$; if $g_\perp >
\left( g_\up+g_\dn \right)/2$ the system will flow to $r=0$,
$|r_\up'|=\sqrt{v_\up/v_\dn}$.


\section{Discussion and summary}
\label{theend}

Estimates of the $g$ couplings have been made for the Hubbard
model in a magnetic field\cite{penk1993}, where the inequality
$g_{2\perp} > \frac 1 2 \left(g_{2\up} + g_{2\dn}\right)$ was
obtained in the strong coupling regime ($U>t$). Using the results
for the renormalized Fermi velocities\cite{penk1993} it is seen
that a regime with $g_\perp > \frac 1 2 \left(g_\up +
g_\dn\right)$ is possible, as well as the opposite inequality.

In the spin-flip reflector fixed point, the spinor wave function
for an incident spin majority electron from the left  has the
asymptotic behavior:
\begin{eqnarray}
\psi(z<0) \simeq \left(
\begin{array}{c}  e^{ik_\up z} \\   r_\up'  e^{-ik_\dn z} \end{array}\right)\,,
\end{eqnarray}
for which the spatial distribution of the spin vector obeys
\begin{eqnarray}
r_\up'  e^{-i\left(k_\up+ k_\dn \right)z} =
e^{i\phi(z)}\tan\frac{\theta(z)}{2} \,,
\end{eqnarray}
where $\phi$ and $\theta$ are the spherical angles of the spin
vector \cite{auerbach}. Since $r_\up'$ has modulus 1 and a finite
phase, then  $\theta(z)=\pi/2$ implying that the conduction
electron spin is normal to the wire and winds around it in
"circularly polarized" fashion with azimuthal angle
$\phi(z)=-\left(k_\up+ k_\dn\right)z$ for $z<0$. For  $z>0$,
$\phi(z)$ changes sign. Thus, the resulting spin density Friedel
oscillations in the Luttinger liquid have such a structure and
decay with distance as a power law.

Our RG method, being perturbative in the electron interactions, 
is valid when the latter are weak. 
But it allows for the scattering barrier 
to be treated exactly, whatever its strength, and hence predict the 
fixed point to which a particular barrier will flow once 
the interactions are turned on.


In summary, we considered the magneto-transport properties a
magnetized one dimensional Luttinger liquid containing  a
localized disorder of the magnetization. We addressed
two cases depending on the origin of this disorder: 1) a stray
magnetic field of an STM tip disturbing the magnetization beneath it,
or 2)   a transverse domain wall (DW). Our theoretical analysis
relies on a renormalization group treatment of the electron
interactions that provides us with scaling equations for the
transmission and reflection amplitudes as the bandwidth is
progressively reduced. We find  two  $T=0$ insulator fixed points:
one in which the carriers are reflected in the same spin channel
and another where the carrier's spin is reversed upon reflection
leading to  a finite spin current without a charge current.  We
also found the role of pure potential scattering in driving 
the DW to the spin-flip insulator phase when the transverse field of the DW
is arbitrarily weak.


\section*{Acknowledgments}
Discussions with P.D. Sacramento are gratefully acknowledged. This
research was supported by Portuguese program POCI under Grant
No. POCI/FIS/58746/2004, German DFG under Grant No. SSP 1165, BE216/3-2,
Polish Ministry of Science and Higher Education as a research project in
2006--2009, and the STCU Grant No. 3098 in Ukraine. M.A.N.A. would like
to thank the hospitality of the Max-Planck Institut f\"ur
Mikrostrukturphysik, Halle (Germany).

\appendix





\begin{thebibliography}{99}
\bibitem{comment}
For a thick
domain wall the electron motion is diffusive inside the wall and
the wall's contribution to  the resistance may even turn negative
because of the single-particle classical (Drude) or
quantum (weak localization) effects.
\end{thebibliography}
\end{document}